\documentstyle[12pt,psfig]{article}
\newcommand{\be}{\begin{equation}}
\newcommand{\ee}{\end{equation}}
\newcommand{\bea}{\begin{eqnarray}}
\newcommand{\eea}{\end{eqnarray}}
\newcommand{\PD}{{\partial}}

\begin{document}
\begin{flushright}
UFTP--435/1997\\
Duke-TH-97-140\\
\end{flushright}
\vspace*{2cm}
\begin{center}
{\Large\bf Nonequilibrium fluid-dynamics in the early stage\\ 
of ultrarelativistic heavy-ion collisions}
\\[2cm]
{\bf J.\ Brachmann, A.\ Dumitru,
J.A.\ Maruhn, H.\ St\"ocker, W.\ Greiner}
\\[0.2cm]
{\small Institut f\"ur Theoretische Physik der J.W. Goethe-Universit\"at}\\
{\small Postfach 111932, D-60054 Frankfurt a.M., Germany}
\\[0.8cm]
{\bf D.H.\ Rischke}
\\[0.2cm]
{\small Department of Physics, Duke University}\\
{\small Durham, NC 27708-0305, U.S.A.}
\\[1cm]
{\large March 1997}
\end{center}
\vspace*{2cm}
\newpage
\begin{abstract}
To describe ultrarelativistic heavy-ion collisions we construct
a three-fluid hydrodynamical model. In contrast to one-fluid hydrodynamics,
it accounts for the finite stopping power of nuclear matter, 
i.e.\ for nonequilibrium effects in the early stage of the
reaction. Within this model, we study baryon dynamics in the
BNL-AGS energy range. For the system Au+Au we find that kinetic
equilibrium between projectile and target nucleons is established only
after a time $t_{CM}^{eq}\approx 5~fm/c\simeq 2R_{Au}/\gamma_{CM}$. 
Observables which 
are sensitive to the early stage of the collision (like e.g.\ nucleon flow)
therefore differ considerably from those calculated in the one-fluid model.
\end{abstract}
\newpage

\section{Introduction}
One of the central interests of heavy-ion physics is the study of the
equation of state (EoS) of hot and highly compressed nuclear
matter~\cite{StoePR}. The success of the
hydrodynamical model~\cite{LLhydro,Clare} in describing
the gross properties of heavy ion reactions at BEVALAC energies
indicated that a 
collective and equilibrated state of highly compressed nuclear matter
is indeed created in collisions of heavy nuclei. 
In particular, experiments (qualitatively) confirmed~\cite{BEVALAC}
\begin{enumerate}
\item enhanced particle emission into preferred solid angles in collisions of a 
heavy target penetrated by a light projectile with velocity larger than 
the speed of sound in nuclear matter, indicating the occurrence of a mach
cone~\cite{Schock,StoePRL},
\item collective flow of matter in the reaction plane ({\em
bounce-off}) and perpendicular to the reaction plane
({\em squeeze-out}) in semiperipheral collisions of
heavy nuclei~\cite{StoePRL,Buchw},
\item the linear increase of the pion multiplicity with beam energy due to
entropy production. In this model entropy is produced in the compression
shock front, the subsequent expansion
proceeds via simple waves which -- in the ideal fluid limit -- 
conserve entropy and thus the pion multiplicity~\cite{linPith}.
\end{enumerate}

These successes motivated the application of the (relativistic) hydrodynamical
model to higher-energy heavy-ion reactions.
At very high baryon densities and / or temperatures a phase transition
from ordinary hadronic matter to a quark-gluon plasma (QGP) is
expected~\cite{StoePR,QGPtrans}. In the (one-fluid) hydrodynamical model the
energy and baryon densities necessary for this phase transition are already
reached in the BNL-AGS energy regime~\cite{Yaris}. Within this model,
the effects of a (first order-like) phase transition lead to
\begin{enumerate}
\item a plateau in the excitation function of $\langle p_T
\rangle$~\cite{vanHove},
\item a plateau in the excitation function of the pion
multiplicity~\cite{Bugaev},
\item a local minimum in the excitation function of the collective
nucleon flow $\langle p_x^{dir}/N\rangle$~{\cite{otherflow,Yaris}},
\item a prolonged lifetime of the system at the ``softest point'' of the
EoS~\cite{Hung}.
\end{enumerate}

However, the question arises whether this model is still valid at
ultrarelativistic energies (AGS and above), where these effects are predicted.
In the one-fluid model instantaneous thermalization of projectile, target, and
produced particles is assumed. Therefore, momentum conservation leads to
immediate complete stopping at the contact surface (the nuclei do not
penetrate each other) and then energy conservation requires that all the
kinetic energy is converted into internal excitation of the stopped matter.
As a consequence, huge (midrapidity) energy densities, temperatures,
and baryon densities are produced. Thus, the conditions for the phase
transition are already reached at comparatively low bombarding energy.

In this paper we extend the hydrodynamical model in such a way that
thermalization of projectile and target does not happen instantaneously.
Within our model, projectile and target decelerate gradually while
penetrating each other. We study the influence of these nonequilibrium
effects on baryon stopping, directed baryon flow, and the compression of
the nuclei. Based on these results, we comment on the validity of the
one-fluid hydrodynamical model at ultrarelativistic energies and our
expectations concerning the above mentioned predictions of this model.

\section{The three-fluid model}
In this section we present the physical motivation for introducing three 
fluids and discuss in detail the interaction sourceterms (see also
refs.\ \cite{2fluid1, 2fluid2, 2fluid3}).

\subsection{Physical motivation}
\label{motiv}
The description of the early stage of a heavy-ion collision within the
framework of ideal one-fluid hydrodynamics tends to become more and more 
unrealistic, when nuclei collide at AGS-energies ($\sqrt{s} \approx 5AGeV$)
and above. The major reason for that is that the assumption
of {\em instantaneous} local thermalization of projectile, target and
produced particles fails at ultrarelativistic bombarding energies. 
This is due to the finite stopping power of nuclear matter. 
As a consequence, one-fluid hydrodynamics overestimates the baryon and
energy densities at midrapidity during the early stage of the reaction.
This in turn yields a high pressure in the
central region, which produces a large sideward flow in the reaction plane 
(the so-called {\em bounce-off}) \cite{Yaris}.

To get a rough estimate of the compression at midrapidity, one may employ the
one-dimensional shock model \cite{Schock1} with an ultrarelativistic
ideal gas EoS ($p=\epsilon/3$) for the shocked matter. The baryon and
energy densities are then given by
\bea
n/n_0 & = & 4\gamma_{cm} \,-\, 3/\gamma_{cm} \quad, \\
\epsilon/\epsilon_0 & = & 4\gamma_{cm}^2 \,-\, 3 \quad. \nonumber
\eea
$n_0\simeq0.16~fm^{-3}$ and $\epsilon_0\simeq0.15~GeV/fm^{3}$ denote ground
state baryon
and energy density of nuclear matter. The linear increase of the central
baryon density with $\gamma_{cm}\propto\sqrt{s}$ becomes
questionable for $\gamma_{cm}\gg1$, since the assumption of instantaneous 
thermalization of projectile and target matter ceases to be valid.

To account for nonequilibrium effects in the early stage of the
reaction we introduce several fluids in our model.
Considering collisions of single protons at that high energies, one observes
a strongly forward-backward--peaked rapidity distribution 
\cite{refppdNdy}.
Each proton loses about one unit of rapidity and is deflected only by a
small angle. The projectile and the target protons remain
separated in phase space, if the incident energy is high enough, and
can be identified as the projectile respectively target proton
even after the collision. This means that to first approximation 
the two baryon flows (projectile and target) do not exchange baryon charge. 
For this reason we consider two fluids that correspond to projectile (fluid 1) 
and target nucleons (fluid 2).
Due to inelastic nucleon-nucleon scattering, energy in form of newly
produced particles is deposited at midrapidity. Since the two nucleonic
fluids do not populate midrapidity in the beginning of the reaction,
if the initial rapidity gap was large enough,
the newly created particles are also separated in phase space. This
motivates to collect them in a third fluid -- the {\em fireball}.

In our approach projectile and target fluids penetrate each other
and decelerate gradually, not instantaneously.
Due to inelastic $NN$ scattering the fireball is dynamically produced 
during the compressional phase of the reaction. Since the time scale of
energy deposition is rather short, we neglect 
the rescattering of the nucleonic fluids with the fireball 
during this early stage. Later, when all three fluids
populate the same phase space region and thermalize, they should
be merged into one. Then,
hydrodynamical expansion becomes the dominant process.

\subsection{Multi-fluid relativistic ideal hydrodynamics}
\label{multifluidhydro}
The basic equations of relativistic one-fluid hydrodynamics are the
conservation of energy and momentum,
\be\label{emconserv}
\partial_\mu T^{\mu\nu} = 0\quad.
\ee
If we assume an ideal fluid (i.e., neglecting dissipative effects),
the energy-momentum tensor $T^{\mu\nu}$ reads ~\cite{LLhydro}
\be
T^{\mu\nu}=(\epsilon+p) u^\mu u^\nu -p g^{\mu\nu} \quad,
\ee
where $\epsilon(x)$ is the energy density in the local rest frame,
$p(x)$ the pressure, $u^\mu(x)=\gamma(1,\vec{v})$ the 4-velocity field
(normalized to unity, $u^\mu u_\mu=1$), and $g^{\mu\nu}={\rm diag}
(1,-1,-1,-1)$ the metric tensor.

In addition, there may exist several conserved currents $j_i^\mu$, for which
the continuity equations
\be \label{bconserv}
\partial_\mu j_i^\mu = 0
\ee
hold. We will consider only one such conserved current, namely the net baryon
current
\be
\label{current}
j^\mu=n u^\mu \quad,
\ee
where $n(x)$ denotes the net baryon density in the local rest frame.

As computational frame we choose the frame where the longitudinal 
velocities of projectile and target nuclei are of equal magnitude. 
In this frame equations~(\ref{emconserv},\ref{bconserv}) read
\bea \label{HEq2a}
\PD_t E + \PD_i (Ev^i) & = & -\PD_i(p{v^i})\quad,\\
\PD_t M^j + \PD_i (M^jv^i)& = & -\PD_j p\quad,\nonumber\\
\PD_t R + \PD_i (Rv^i) & = & 0      \quad.\nonumber
\eea
$E$, $\vec{M}$, $R$ denote energy, momentum, and baryon density in the
computational frame. 
They are related to the local rest frame quantities $\epsilon$, $p$
and $n$ via
\begin{eqnarray} \label{untangle}
E &\equiv & T^{00} = \gamma^2(\epsilon+p)-p\quad,\\
M^i &\equiv & T^{i0} = \gamma^2(\epsilon+p)v^i\quad,\nonumber\\
R &\equiv & j^{0} = \gamma n\quad.\nonumber
\end{eqnarray}
The system~(\ref{HEq2a},\ref{untangle}) of coupled differential equations is
closed via the assumption of local thermodynamic equilibrium and by
specifying an EoS in the form $p(\epsilon,n)$.

In hydrodynamics for $N$ fluids, we split the energy-momentum tensor and
the baryon current of the total system into a sum of $N$ terms,
\begin{eqnarray} \label{flsubdiv}
T^{\mu\nu} &=& T_1^{\mu\nu}+T_2^{\mu\nu}+\cdots +T_N^{\mu\nu}\quad,\\
j^\mu &=& j_1^\mu+j_2^\mu+\cdots +j_N^\mu\quad.\nonumber
\end{eqnarray}
The total system is now subdivided into $N$ individual fluids.
Note that eqs.\ (\ref{flsubdiv}) hold {\em locally}, i.e.\ different
fluids may coexist at the same space-time point.
Each individual energy-momentum tensor 
$T_l^{\mu\nu}$ and baryon current $j_l^\mu$ does not need to be conserved,
since the $N$ subsystems may exchange energy, momentum and baryon charge:
\begin{eqnarray}\label{HEq2b}
\partial_\mu T_l^{\mu\nu} &=& F_l^\nu \quad, \\
\partial_\mu j_l^\mu &=& S_l \quad\quad (l=1,\cdot,N)\quad. \nonumber
\end{eqnarray}
$F_l^\nu$ denote energy and momentum sources of fluid $l$, and $S_l$
the baryon charge exchange for each fluid.
Due to eqs.~(\ref{emconserv},\ref{bconserv}), i.e.\ the conservation of
total energy, momentum and baryon charge, we have the additional equations
\begin{eqnarray} \label{sourcessum}
F_1^\nu+ F_2^\nu+\cdots+ F_N^\nu &=& 0 \quad, \\
S_1+S_2+\cdots+ S_N &=& 0 \quad. \nonumber
\end{eqnarray}
Following the discussion in section~\ref{motiv}, we choose $N=3$ and
identify fluid $1$ with the projectile and fluid $2$ with the target nucleons.
Fluid $3$ collects the energy loss
of projectile and target nucleons due to inelastic collisions, and
will thus be identified with the produced particles.
At time $t=0$ (in the computational frame), the third fluid is
not yet existent: $\quad j_3^\mu(t=0,\vec{x})=0$, 
\quad $T_3^{\mu\nu}(t=0,\vec{x})=0$.

\subsection{Coupling terms between the fluids}
We now discuss the couplings between the three fluids, based on the
assumption that they are well separated in rapidity.

We first discuss the interaction between projectile and target.
The total one-particle distribution function can be written as the sum
of two individual distribution functions for projectile and target nucleons,
respectively:
\be
f(t,\vec{x},\vec{p}) = f_P(t,\vec{x},\vec{p}) + f_T(t,\vec{x},\vec{p}) \quad .
\ee

Due to the strongly forward-backward peaked cross section
(see discussion in section \ref{motiv}) in
high energy $pp$-collisions it is possible to identify the projectile
respectively target nucleon even after the reaction.
The nucleon scattered into the forward hemisphere 
belongs by definition to the same flow as the one previously
heading in this direction.
Consequently, projectile and target fluid do not exchange baryon charge:
$S_1=S_2=0$. From eq.~(\ref{sourcessum}) and the fact that the fireball 
is also separated in phase space from the nucleonic fluids, it follows that 
the fireball remains net baryon free, $S_3=0$.

The source terms of the energy-momentum tensor $F_l^\nu$ in eq.~(\ref{HEq2b}) 
describe the energy respectively momentum loss of fluid $l$ per volume
and per unit time. We define
\be \label{sourcesform}
F_l^\nu = n_1\, n_2\, \left< v_{M{\o}ller}\, \int\limits_{\tilde{p}_{||}'>0}
\!{\rm d}\sigma_{{\rm NN}\rightarrow{\rm NX}}\, (p'-p)^\nu \right> \quad , 
\quad {\rm d}\sigma_{{\rm NN}\rightarrow{\rm NX}} = 
\sigma_{{\rm NN}\rightarrow{\rm NX}}^{inv} \, \frac{{\rm d}^3p'}{E'} \quad.
\ee
This is simply the collision rate per volume 
$n_1\, n_2\, v_{M{\o}ller}\, {\rm d}\sigma$
for scattering into the invariant phase space volume ${\rm d}^3p'/E'$ 
times the average 4-momentum loss $(p'-p)^\nu$ in a single binary 
$NN$-collision. 
$\sigma_{{\rm NN}\rightarrow{\rm NX}}^{inv}$ denotes the invariant differential
cross section. 
$p'^\nu$, $p^\nu$ are the 4-momenta of the particle after 
and before the collision, respectively. 
The integration extends over the forward hemisphere of the cross section only
($\tilde{p}_{||}'>0$). Quantities with a tilde refer to the center-of-mass
frame of
the binary $NN$ collision. The $n_i$ denote the local rest frame baryon
densities of projectile and target, as defined in
eqs.\ (\ref{current}) and (\ref{untangle}).
$v_{M{\o}ller}$ denotes the invariant relative velocity of 
the two fluids~\cite{MoellerVel}:
\be \label{moeller}
v_{M{\o}ller} = \sqrt{\left( u_1^\mu \,u_{2\mu} \right)^2 - 1}
\quad.
\ee
The $\langle\cdot\rangle$-averaging over the individual particle distribution
functions provides a smearing of the sourceterms (accounting for the thermal
and Fermi momenta of the particles), especially of the
M{\o}ller velocity. It is negligible during the compressional stage of the
reaction since the internal velocities within each fluid element are small 
compared to the M{\o}ller velocity of the colliding fluids. 
We therefore ignore this averaging. Instead,
when thermal and relative velocities of the colliding fluids become
comparable in the later stage of the collision, the fluids are unified
(see section \ref{onef}).

In principle it is always possible to split the sourceterms in a
symmetric and an antisymmetric part with respect to the fluid indices
($1\leftrightarrow2$):
\bea
\label{sourcesprinciple}
\partial_\mu T_1^{\mu\nu} & = & f_{exchange}^\nu -f_{loss}^\nu\quad, \nonumber\\
\partial_\mu T_2^{\mu\nu} & = & -f_{exchange}^\nu -f_{loss}^\nu\quad, \\
\partial_\mu T_3^{\mu\nu} & = & 2f_{loss}^\nu\quad.\nonumber
\eea
The antisymmetric term $f_{exchange}^\nu$ describes the exchange of energy and 
momentum between projectile and target fluid, while $f_{loss}^\nu$ denotes the 
loss of energy and momentum transferred to the fireball.

Following \cite{2fluid1,2fluid2,2fluid3} we split the integral
in eq.~(\ref{sourcesform}) into two moments,
\bea\label{F1source}
F_1^\nu & = & \frac{1}{2}\, n_1\, n_2\, v_{M{\o}ller}\,
[(p_2-p_1)^\nu \sigma_P(s)\,-\,(p_2+p_1)^\nu \sigma_E(s)]\quad,\\
F_2^\nu & = & \frac{1}{2}\, n_2\, n_1\, v_{M{\o}ller}\,
[(p_1-p_2)^\nu \sigma_P(s)\,-\,(p_1+p_2)^\nu \sigma_E(s)] \nonumber \quad .
\eea
Since we neglect the $\langle\cdot\rangle$-averaging, the particle momenta are 
simply given by $p_{1,2}^\nu = m_N u_{1,2}^\nu$.
The moments of the cross section $\sigma_E$ and $\sigma_P$ are defined as
\be \label{moments}
\sigma_E(s) = \int\limits_{\tilde{p}_{||}'>0}\!{\rm d}\sigma_{{\rm
NN}\rightarrow{\rm NX}}\,\left(1-\frac{\tilde{E}'}
{\tilde{E}}\right) \quad, \quad
\sigma_P(s) = \int\limits_{\tilde{p}_{||}'>0}\!{\rm d}\sigma_{{\rm
NN}\rightarrow{\rm NX}}\,\left(1-\frac{\tilde{p}'_{||}}
{\tilde{p}_{||}}\right)
\ee
and parametrize the mean energy respectively longitudinal 
momentum loss in a single nucleon-nucleon collision.
$E$, $\vec{p}$ are energy and momentum of the nucleon before
the collision, while $E'$, $\vec{p'}$ denote the same quantity after the
collision.
Again, the integration is carried out only over the forward hemisphere of the
cross section ($\tilde{p}'_{||}>0$).

Comparing eqs.~(\ref{sourcesprinciple}) and eqs.~(\ref{F1source})
we see that $\sigma_E$ corresponds to the symmetric term and leads to the 
production of the fireball. Analogously, $\sigma_P$ is the antisymmetric 
term describing the exchange of energy and momentum between
projectile and target fluids. Also, from this form of the coupling terms
it immediately follows that the third fluid is produced at midrapidity:
in the CMS $\vec{p_1} = -\vec{p_2}$ and thus $\vec{F_3} =\vec{0}$.  
The hydrodynamical eqs.~(\ref{sourcessum}) and the specific form of the
sourceterms, eq.~(\ref{F1source}), can also be derived from the Boltzmann 
equation \cite{2fluid3}.

We employ the parametrizations of $\sigma_{P,E}$ from the two-fluid model 
of refs.\ \cite{2fluid1,2fluid2,2fluid3}.
There the cross sections of free binary $NN$ collisions ($NN \rightarrow N^*X$) 
have been used to evaluate the moments $\sigma_{P,E}$.
The incident baryons are assumed to be nucleons, whereas in the final
state we sum over all baryons $N^*$.
The rescattering with the fireball ($N\pi \rightarrow X$)
is
not considered, since the contribution to the total cross section
outside the $\Delta$-resonance is small 
at high bombarding energies. Since we focus on the dynamics of the nucleons
in this paper, we do not expect this to play an important role.

\subsection{Equations of state}
To close the coupled system of differential equations~(\ref{HEq2b})
we assume local thermodynamic equilibrium within each fluid (but not
in the total system~!) and an EoS for each fluid.

For the target and projectile fluids we employ an EoS of an ideal nucleon gas 
plus compression energy:
\be \label{idnucleos}
p = \zeta (\epsilon-E_c n) +p_c \quad.
\ee
$E_c$ is the compression energy per nucleon and $p_c$ the compressional
pressure. Since the nucleon mass $m_N$ is large compared to the typical
temperatures of the nucleons, we choose the nonrelativistic limit
$\zeta=2/3$. For the compression energy, we employ the ansatz
\be\label{comprene}
E_c=\frac{k_c}{18n n_0} (n-n_0)^2 +m_N +W_0 \quad,
\ee
i.e., $E_c$ increases linearly with the baryon density $n$ for $n\gg n_0
\approx0.16~fm^{-3}$. We assume a ground-state (in-)compressibility 
$k_c=300~MeV$ and a binding energy of $W_0=-16~MeV$. The compressional
pressure $p_c$ and the velocity of sound $c_S$ at temperature $T=0$ are given by
\bea
p_c &=& -\frac{dE_c}{dn^{-1}}=n^2\frac{dE_c}{dn}=\frac{k_c}{18n_0}
(n^2-n_0^2) \quad,\\
c_S^2 &=& \frac{dp}{d\epsilon}=\frac{n/n_0}
{n/n_0-1+9(m_N+W_0)/k_c}\quad.\nonumber
\eea
Ansatz~(\ref{comprene}) thus ensures $c_S(n,T=0)<1$ as long as
$k_c<9(m_N+W_0)\approx8.3~GeV$.

For the third fluid we employ an EoS of an ideal gas of $\pi$, $\eta$,
$\rho$, $\omega$ mesons with a phase transition to QGP at $T_c = 160~MeV$
\cite{Dum95a}. However, here we do not discuss the dynamics of the
third fluid.

\subsection{One-fluid transition}
\label{onef}
In the later stage of the collision the nucleonic fluids stop.
Their relative velocity (M{\o}ller velocity)
is then comparable to the internal thermal velocities.
The two fluids are no longer separated in phase space, so that the main
assumption for a two-fluid region does not hold anymore.
Moreover, the coupling sourceterms, eqs.~(\ref{F1source}), cease to be valid, 
since they do not account for thermal smearing and vanish linearly with
the M{\o}ller velocity.
Thus, two infinitesimal fluid elements with equal fluid velocities, $u_1^\mu 
= u_2^\mu$, would not exchange energy and momentum, even if their
temperatures were different.
Because the particle velocities are thermally distributed
around the fluid velocity, there should be a non-zero coupling that further
drives the two fluids into common local thermal equilibrium.
Since we do not account for this, the two fluids are merged, if
the M{\o}ller velocity (\ref{moeller}) is comparable to the 
root-mean-square velocity in a nonrelativistic degenerate Fermi gas
or a nonrelativistic Boltzmann gas:
\bea
\label{boltzmangas}
\left(\Delta v\right)^2_{Boltzmann} & = & 
\max_{l=1,2} \left\{ 3 T_l/m_N \right\} \quad,\\
\label{fermigas}
\left(\Delta v\right)^2_{Fermi} & = & 
\max_{l=1,2} \left\{ 2 \epsilon_F(n_l)/m_N \right\} \quad,
\eea
where
\be
\epsilon_F(n_l) =  \left(\frac{3}{2}\pi^2n_l\right)^{2/3}/2m_N 
\ee
denotes the Fermi energy.
In contrast to \cite{Amsden,Clare,2fluid1}, the transition is done abruptly
without a smooth transition from two to one fluid.

Under the constraining assumption that both fluids flow with a common 
velocity $\vec{v}$ after unification,
the set of eqs.~(\ref{untangle}) is solved for the local rest frame densities 
$\epsilon$ and $n$ by inserting the total densities $E=E_1+E_2$, $R=R_1+R_2$,
$\vec{M}=\vec{M}_1+\vec{M}_2$. 
By virtue of the EoS, the total pressure $p=p(\epsilon,n)$ is also known.
Additive quantities, i.e.\ the pressure $p=p_1+p_2$, the baryon density
$n = n_1 + n_2$ and the energy density $\epsilon=\epsilon_1+\epsilon_2$ are 
then redistributed to the individual fluids according to the ratio of the
baryon densities in the computational frame~\cite{Amsden,Clare}:
\be \label{restribution}
p_i=p\,\frac{R_i}{R_1+R_2} \quad,\quad
n_i=n\,\frac{R_i}{R_1+R_2} \quad,\quad
\epsilon_i=\epsilon\,\frac{R_i}{R_1+R_2} \quad,\qquad i\in\{1,2\}\quad.
\ee
Since this procedure changes the individual fluid velocities $v_i$ 
to the common velocity $v$,
the propagated quantities $R_i$, $E_i$, $\vec{M_i}$ have to be 
redetermined from $n_i$, $\epsilon_i$ and $p_i$
according to equations (\ref{untangle}) with the common velocity
$v$ and the corresponding $\gamma$ factor. 
Let $\tilde{R}_i$, $\tilde{E}_i$, $\vec{\tilde{M}_i}$ be the 
resulting quantities.
Our special choice of the redistribution, eq.~(\ref{restribution}),
implies that the individual baryon densities in the computational frame
are not changed: $\tilde{R}_i=R_i\quad.$ 

It can be easily proven that the total quantities
$\tilde{E}_1+\tilde{E}_2=E=E_1+E_2$, $\tilde{R}_1+\tilde{R}_2=R=R_1+R_2$,
$\tilde{\vec{M}}_1+\vec{\tilde{M}}_2=\vec{M}=\vec{M}_1+\vec{M}_2$
fulfill the one-fluid hydrodynamical eqs.~(\ref{HEq2a}).
That makes it possible to propagate the fluids separately. Even
if that has no physical meaning for the individual fluid, one can tell 
what amounts in phase space originated from projectile or target, respectively.

Moreover we will use this feature to simulate a one-fluid model with our 
three-fluid model, by switching off the production of the fireball
(setting $\sigma_E=0$) and forcing the local unification of the 
two nucleonic fluids at any instant.
This is what we will call the {\em one-fluid limit} of the three-fluid model.
Formally, the one-fluid limit corresponds to $\Delta v=\infty$ in
eqs.~(\ref{boltzmangas},\ref{fermigas}).

In the present stage of model development, only the one-fluid 
transition of the {\em nucleonic} fluids is implemented. The unification of
{\em all three}
fluids is recommended and in preparation because much of the
nucleonic fluids finally end up at midrapidity, where the third fluid is
produced. This would slightly increase the pressure in the central region
and enhance the directed and radial flow of the nucleons.

\subsection{Technical realization of the model}
The hydrodynamical eqs.\ (\ref{HEq2a}) are solved on an Eulerian grid
using the SHASTA ({\bf SH}arp {\bf A}nd {\bf S}mooth {\bf T}ransport 
{\bf A}lgorithm) \cite{Bor73,Bor75,Bor76}. This algorithm was extensively 
tested in \cite{Dirk1,Dirk2}.
Each of the three fluids is propagated on its own grid with eight different 
{\em densities} ($R$,$\vec{M}$,$E$,$n$,$\epsilon$,$p$).
For each fluid the three-dimensional propagation
is performed using the method of time-step splitting (operator splitting)
\cite{Dirk1}, i.e.\ the 3-divergence operators on both sides of
eqs.\ (\ref{HEq2a}) are {\em not} summed over the contracting index $i$.
One solves sequentially equations of the form:
\be
\PD_t U + \PD_i (Uv_i+f) = 0\quad,\quad \mbox{for $i=x,y,z$} \quad. 
\ee
Here $U$ stands for either $R,M_i$, or $E$ and $-\PD_i f$ denotes the
corresponding right-hand side of eqs.\ (\ref{HEq2a}), e.g.\ in case of
$U=E$ and $i=x$ we have $-\PD_x f= -\PD_x(p v_x)$.
The $z$(beam)-direction is propagated first, those $z$-propagated 
densities are then propagated in $x$-direction (impact parameter) and 
finally propagated in $y$-direction (out-of-plane). 
This order is kept fixed for reasons of sim\-pli\-ci\-ty.
By permutation the symmetry of the propagation could be enhanced \cite{Dirk1}.
Obviously the {\em propagational sourceterms} $-\PD_i f$ are included in the
propagation.

In contrast, we exclude in our calculations the {\em coupling sourceterms} 
(right-hand side of eqs.\ (\ref{HEq2b})) from the propagation.
In our implementation the fluids are first propagated without coupling terms
and their local rest frame densities are calculated according to 
eqs.~(\ref{untangle}).
From these densities the coupling sourceterms
(eqs.\ (\ref{F1source})) are calculated and added.
Since this coupling changes the energy and momentum 
density of the individual fluids, eqs.~(\ref{untangle}) have to be solved
again in order to accomplish the full time step. 
Applying this procedure it is easy to control the effect of the coupling,
e.g.\ a cell cannot become more than empty by extracting energy from it.
In such a case the coupling would not be applied to any of the fluids.

In order to minimize the Lorentz contraction we choose the equal-velocity frame
as the computational frame. In the equal-velocity frame both projectile
and target nucleus have the same velocity (with opposite sign). 
In the case of two nuclei
of the same mass this is identical to the center-of-mass frame.
Because of the occurrence of acausalities one should not use a
different grid spacing in different spatial directions \cite{Waldhauser}.
Therefore, we use an equidistant grid with grid spacing $\Delta x=\Delta y
=\Delta z=0.15fm$. This choice was found sufficient to provide convergence
in the rapidity distributions.
It ensures the correct number of $NN$ collisions with the appropriate 
colliding energy $\sqrt{s}$.
Our time step is $\Delta t = 0.4\,\Delta x$ to fulfill the
Courant-Friedrichs-Lewy criterion \cite{Courant} for the SHASTA. 

As a consequence of eqs.\ (\ref{untangle}) the cell velocity is:
\be
v = \frac{\left|\vec{M}\right|}{E+p} \quad.
\ee
We call a cell acausal if $E<\left|\vec{M}\right|$.
This happens mainly because in the final step of the SHASTA the
propagated density is distributed uniformly over a cell so that,
on a too coarse grid, some part of the fluid is transported due to this 
smoothing procedure and not by real motion. Another possible origin is that due 
to the coupling of the different fluids too much energy is extracted from a 
fluid compared to the extracted momentum.
We are correcting those acausal cells by rescaling the energy density $E$
with the momentum density:
\be
E_{corrected} = \left( 1+10^{-8}\right) \cdot\left|\vec{M}\right| \quad.
\ee
This corresponds to setting the corrected cells to a maximum velocity $v_{max}$.
By removing the acausal cells instead of correcting them as described above, 
we found for the $\Delta x=0.15fm$ grid used here that less than $5.0\%$ of 
the initial baryon charge and $2.0\%$ of the initial total energy was lost.

\section{Previous models}
Before presenting results we briefly review some earlier work
on which our model is based. The main difference to these
approaches is that in none of the predecessor models the produced particles
were collected in a third fluid and propagated hydrodynamically. They
were two-fluid models only (i.e., $N=2$ in the above equations).

To our knowledge, the first application of multi-fluid hydrodynamics to 
heavy-ion collisions was discussed in ref.~\cite{Amsden}.
Similar to our model, fluids $1$ and $2$ corresponded to the
projectile and target nucleons, respectively. They could exchange energy
and momentum but no baryons. The couplings between the two fluids were
already derived using the underlying concept of binary nucleon-nucleon
interactions. However, only elastic nucleon-nucleon scattering (i.e., no
particle production) was taken into account. In this model
$F_1^\mu+F_2^\mu=0$ and thus the sum of $T_1^{\mu\nu}$
and $T_2^{\mu\nu}$ was conserved.

In ref.\ \cite{Clare} a similar model with a more refined coupling
(more appropriate for the upper BEVALAC energy range, $E^{Kin}_{Lab}
\simeq2~AGeV$) was presented. However, also in this model no midrapidity
fluid was considered, i.e.\ $F^\mu_1+F^\mu_2=0$.

In contrast, the authors of~\cite{Barz} investigated the case
$F_1^\mu+F_2^\mu\neq0$. Thus, the possibility of inelastic reactions in
collisions between projectile and target nucleons was accounted for.
However, the antisymmetric (with respect to exchange of the fluid indices
$1\leftrightarrow2$) term in their couplings (diffusion term),
\be \label{F1plusF2}
F_1^\mu-F_2^\mu= -2Dn_1n_2(u_1^\mu-u_2^\mu)\quad,
\ee
involved a diffusion constant $D$, whereas we employ a $\sqrt{s}$-dependent
$\sigma_P$, cf.\ eq.~(\ref{F1source}). Moreover, in their model the symmetric
term (drain term) was
\be \label{F1minusF2}
F_1^\mu+F_2^\mu= -2q (n_1u_2^\mu+ n_2u_1^\mu) \quad,
\ee
again with a $\sqrt{s}$-independent $q$.
As already mentioned above, the radiative energy losses of fluids $1$ and $2$ 
were not collected in a third fluid and propagated hydrodynamically, in
contrast to our model. Furthermore, the fluids were propagated in 
$(1+1)$-dimensions only, thus calculations of the transverse 
flow of the nucleons were not possible. 
Finally, since merging of the two fluids into one fluid 
(when they reach thermal equilibrium) was also not considered, 
the expansion of the stopped and excited projectile and target fluids 
and collective flow of spectator matter could not be investigated.

In the two-fluid model of \cite{2fluid1,2fluid2,2fluid3}, 
the sourceterms were calculated using experimental data for the elastic
and inelastic $NN$ cross sections and do not contain free parameters like
$D$ and $q$ in eqs.\ (\ref{F1plusF2}) and (\ref{F1minusF2}).
There was no energy loss to the fireball below the pion production threshold 
and the energy deposition in the fireball increases with the c.m.\ energy 
$\sqrt{s}$ of the colliding nucleonic fluids. 
This $\sqrt{s}$ dependence of the sourceterms accounts for the multiple 
production of particles, mainly pions.
However, the authors of \cite{2fluid1,2fluid2,2fluid3} considered only two 
fluids: the energy loss due to the creation of the 
fireball was accounted for in the propagation of projectile/target fluids,
but the fireball itself was not propagated hydrodynamically.
In~\cite{Uli93} the opposite point of view was taken by studying the
production and hydrodynamic expansion of the third fluid in $S+S$ at
$200~AGeV$ (assuming some
global interaction between the nucleons of fluids one and two), however,
disregarding the compression and heating (and thus the hydrodynamic
evolution) of the projectile and target fluids.

In contrast, our three-fluid model propagates all three fluids
hydrodynamically and treats the interactions {\em locally} in space-time.
It thus allows to study spectra of produced 
particles like thermal pions or photons~\cite{Dum95a} as well as
the dynamics of the nucleons.
In this paper we focus on the latter and thus 
neglect the scattering of the fluid of produced particles with the projectile
and target fluids. A more refined investigation of various particle spectra,
as in \cite{BMSW}, is in preparation.

\section{Results}
In this section we present results on compression of projectile respectively
target, baryon stopping and flow. We emphasize the deviations from the
one-fluid limit which is achieved by setting $\sigma_E \equiv 0$ (i.e.\
no energy loss due to particle production) and by merging the projectile
and target fluids at each space-time point from the very beginning.

A remark about the freeze-out should be added here.
Usually a hydrodynamical treatment becomes unrealistic when reaching some
local freeze-out criterion, i.e.\ for example the local temperature drops 
below the freeze-out temperature \cite{tempfreezeout} or the local density 
becomes smaller than a certain freeze-out density. 
Fluid elements would then freeze-out locally
(e.g.\ in boostinvariant hydrodynamics~\cite{Bj} according to their eigentime).
While such a local freeze-out criterion would be a more realistic 
scenario, it is very difficult to realize, especially in a 
fully three-dimensional calculation.

Due to these difficulties and in order not to introduce additional free
parameters (like the freeze-out temperature and / or density), we prefer
to discuss only fluid instead of particle distributions.
Our results (e.g.\ the dN/dY-spectra, cf.\ section~\ref{BarStopp})
are therefore compared to the one-fluid limit, not directly to experimental
data. Also due to this reason, we show the time evolution of the observables
under consideration at fixed center-of-mass time (not at fixed proper time).

\subsection{Density evolution}
\label{denssec}
\begin{figure}[hbp]
\vspace*{-2cm}
\hspace{0cm}{\hbox{\psfig{figure=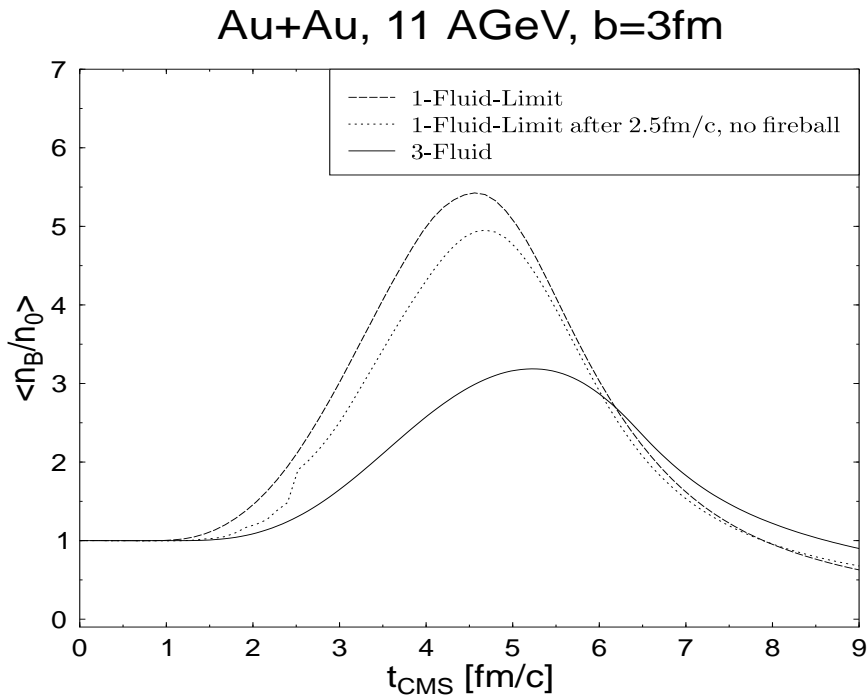,height=15.5cm,width=12.5cm}}}
\vspace*{-4cm}
\caption{Average baryon density of the projectile.}
\label{denspic}

\vspace*{.5cm}
\end{figure}

In section \ref{motiv} we claimed that in a one-fluid model much higher 
densities are reached than in multi-fluid models. 
Figure \ref{denspic} shows the average baryon density of the projectile
nucleus as a function of time. The averaging is defined
as 
\be
\left<n_P\right> = \frac{\int \!{\rm d}^3x \,R_P(x)\, n_P(x)}
{\int \!{\rm d}^3x R_P(x)}\quad.
\ee
where $R_P$ denotes the projectile baryon density in the computational frame, 
cf.\ section~\ref{multifluidhydro}.
Thus it is a measure for the compression of the projectile only.
In the one-fluid limit the maximum projectile density is almost a factor
of two higher and is reached earlier than in the three-fluid model.\\
The dotted line shows a calculation without fireball production ($\sigma_E=0$),
like in the one-fluid limit, but the unification was performed after
$2.5fm/c$.
The evolution first follows more or less the three-fluid curve but yields 
slightly higher densities. 
It was already pointed out in \cite{Barz} that neglecting
the energy loss due to particle production results in higher excitations in
multi-fluid models. At $2.5fm/c$, when instantaneous unification like in the
one-fluid limit is applied, the curve jumps towards the one-fluid limit 
but does not reach the full height
since the nuclei already penetrated partially.

\begin{figure}[hbp]
\hspace{1cm}{\hbox{\psfig{figure=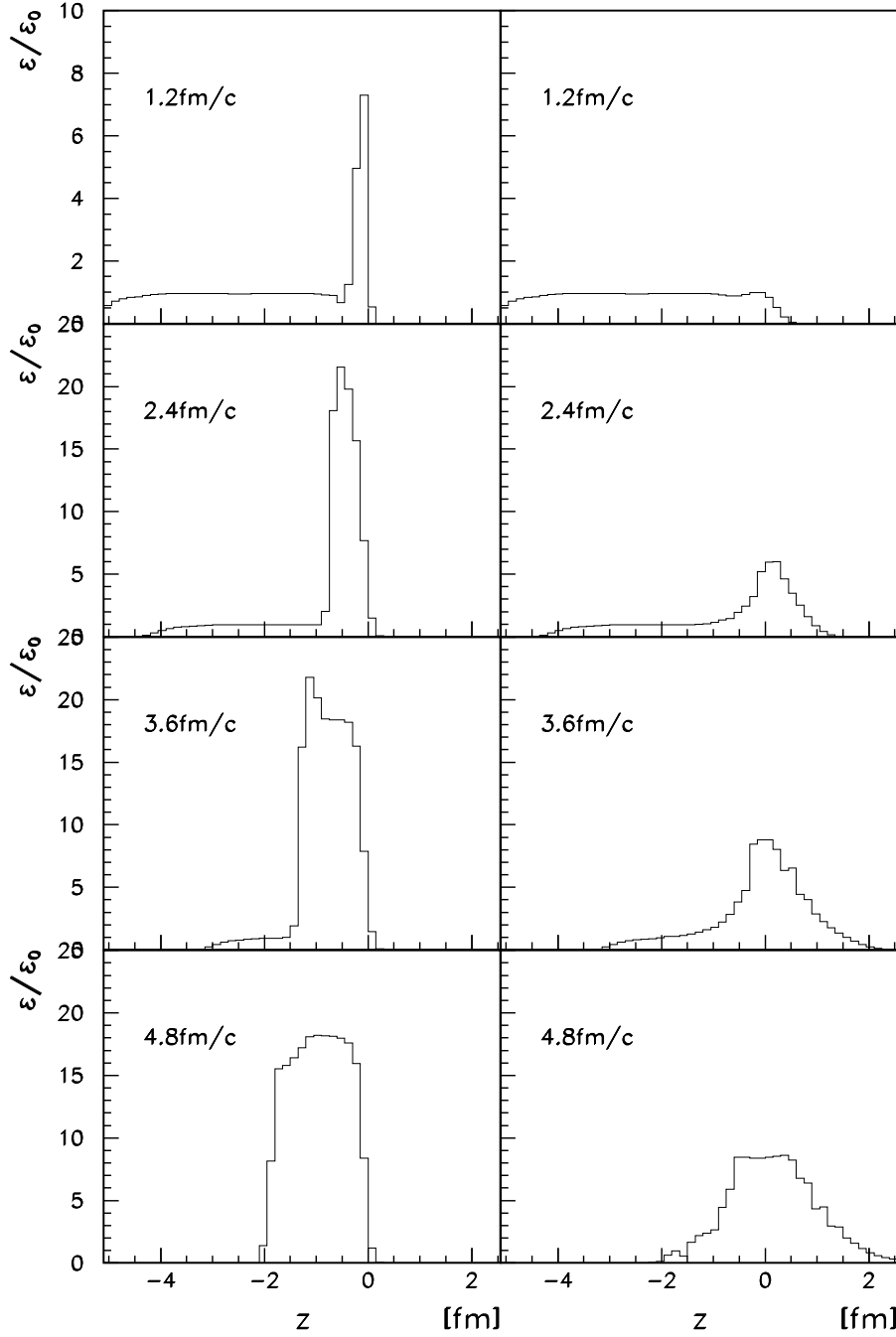,height=20cm,width=13cm}}}
\vspace*{-2cm}
\caption{Energy density profile (in units of $\epsilon_0\approx 0.15~GeV/fm^3$)
of the projectile fluid along the beam axis (in the plane $x=y=0$).}
\label{shockwave}
\vspace*{.5cm}
\end{figure}

The reason for the lower compression of the baryons in the three-fluid model
is the finite stopping length of nucleons in nuclear matter. This is
illustrated in Fig.\ \ref{shockwave} which shows the energy density profile
of the projectile fluid along the beam axis.
In the one-fluid limit the shock wave which moves outwards can be clearly seen,
leaving behind compressed matter with energy density $\epsilon\approx20
\epsilon_0$. Also, no flow through the plane $z=0$ exists
(corresponding to instantaneous, complete stopping of the projectile). 
We checked that the shock-wave velocity and the energy density of the 
shocked matter in this three-dimensional calculation agrees 
satisfactorily well with the one-dimensional shock model.
In contrast,
in the three-fluid model there is no discontinuity in the baryon or
energy density profile which separates the central region from the
incoming ground-state nuclear matter. Due to their non-vanishing
mean-free path some projectile nucleons do pass the ($z=0$)-plane and the
shock front is broadened considerably. As a consequence, the maximum
energy density of the projectile matter does not exceed $10\epsilon_0$.

\subsection{Baryon stopping} \label{BarStopp}
\begin{figure}[hbp]
\hspace{1cm}{\hbox
{\psfig{figure=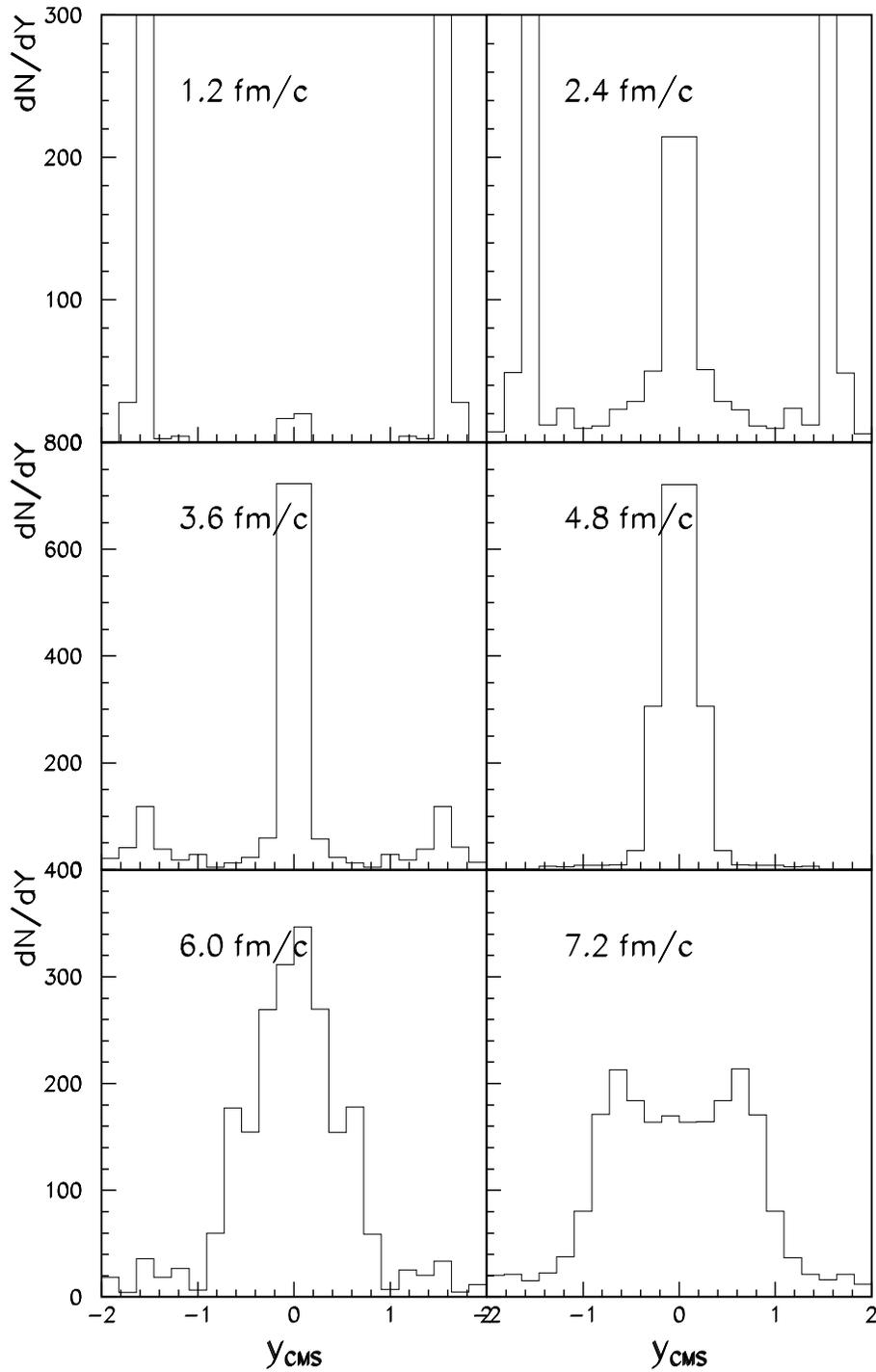,height=20cm,width=13cm}}}
\vspace*{-2cm}
\caption{The evolution of the baryon fluid dN/dY in the one-fluid limit.}
\label{dndy1fluidpic}
\vspace*{.5cm}
\end{figure}

\begin{figure}[hbp]
\hspace{1cm}{\hbox
{\psfig{figure=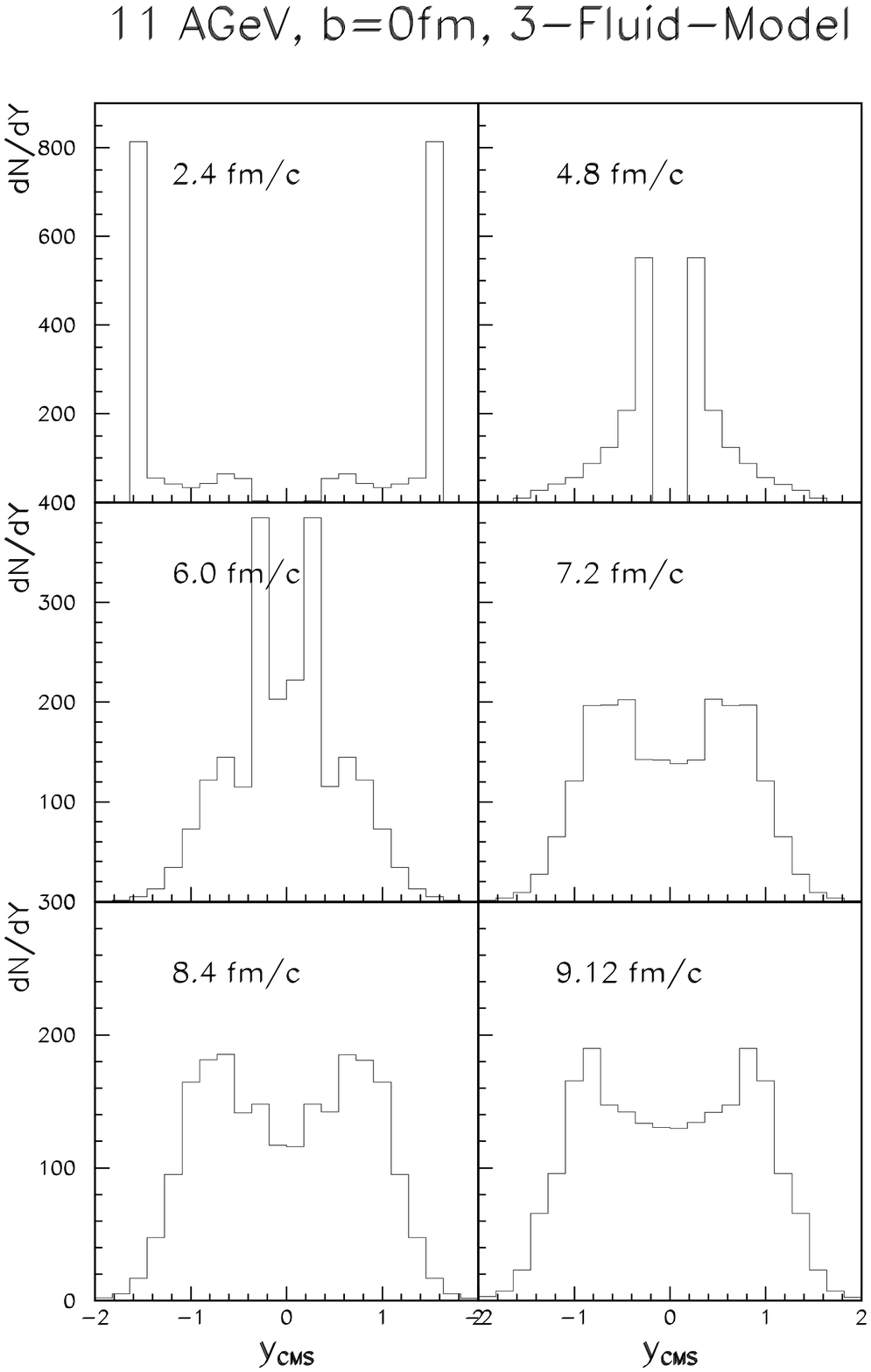,height=20cm,width=13cm}}}
\vspace*{-2cm}
\caption{The evolution of the baryon fluid dN/dY in the three-fluid model.}
\label{dndy3fluidpic}
\vspace*{.5cm}
\end{figure}

Figures \ref{dndy1fluidpic} and \ref{dndy3fluidpic} compare the time evolution 
of the rapidity distributions of the nucleonic fluids 1,2
in the one-fluid limit and the three-fluid model, respectively. 
We again stress that both figures show fluid distributions, i.e.\ the particle
momenta are not thermally smeared.
The time steps in the two figures are not
equal, since the compression reaches its maximum faster in the one-fluid
limit, cf.\ Fig.\ \ref{denspic}.

The major difference between one- and multi-fluid models as claimed in section
\ref{motiv} is becoming obvious: 
At $1.2fm/c$ the fluid rapidity distribution in the one-fluid limit
(Fig.~\ref{dndy1fluidpic}, upper left panel) shows clearly that participant 
matter stops immediately at midrapidity when the nuclei touch, with nothing
in between midrapidity and the projectile or target region,
respectively.
Comparing the rapidity distributions at $4.8fm/c$ (Fig.~\ref{dndy1fluidpic},
middle right panel, and Fig.~\ref{dndy3fluidpic}, upper right panel),
in the one-fluid limit the colliding baryons 
are piling up at midrapidity while in the three-fluid model they are
shifting smoothly towards midrapidity.
Nevertheless, looking at the later stages, i.e.\ $6.0fm/c$, even in the 
three-fluid model the colliding nuclei become more or less stopped 
within about one unit in rapidity before reaccelarating.
Thus, the coupling sourceterms, parametrized by free binary 
nucleon-nucleon collisions, are strong enough to yield a sufficient stopping.

A benefit of our unification procedure is that we can track the
individual nucleonic fluids even after unification since they
are formally propagated on separate grids. We observe 
that projectile and target are changing position in the dN/dY in the one-fluid
limit, while they keep the sign in rapidity in the three-fluid
model, i.e., that projectile and target are reflected in the one-fluid
limit, while heading on in their initial direction in the three-fluid model.

In the late stage of the reaction expansion is becoming the most important
process which is reaccelerating the fluids again in both the one-fluid
limit and the three-fluid model.

\subsection{Equilibration}
\begin{figure}[hbp]
\hspace{1cm}{\hbox
{\psfig{figure=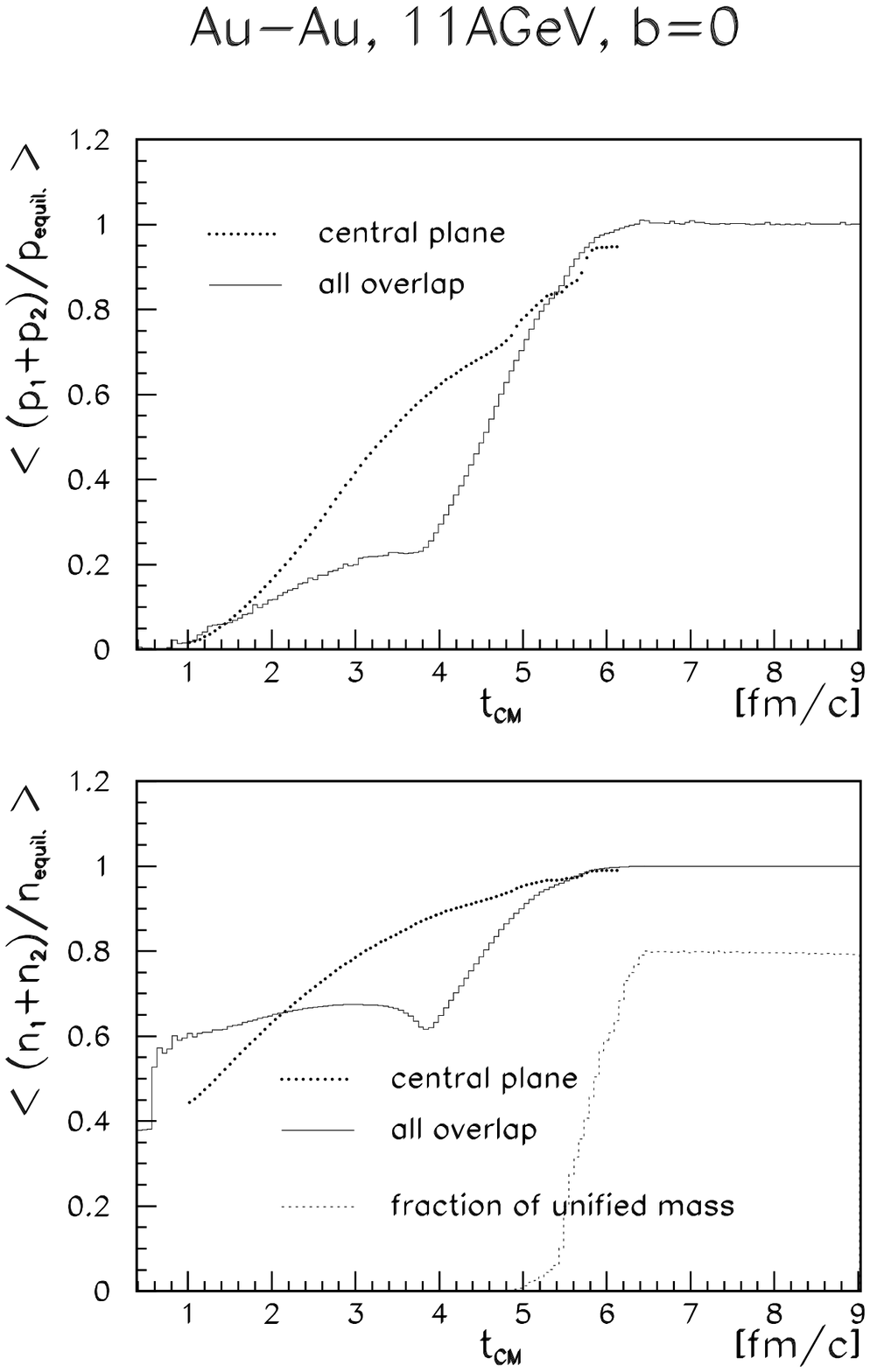,height=20cm,width=13cm}}}
\vspace*{-3cm}
\caption{Evolution of sum of nonequilibrium pressures and rest frame 
baryon densities compared to the corresponding equilibrium values (ratios).}
\label{noneq_eqpic}
\end{figure}

For an equilibrated system the pressure is isotropic.
In contrast, while equilibration is not yet established, the 
transverse and longitudinal pressures (as defined in kinetic
theory~\cite{deGroot}) will differ from each other. In our approach,
this reflects in a deviation of the sum of the partial pressures (of projectile
and target fluids) from the equilibrium pressure.
This is investigated in Fig.\ \ref{noneq_eqpic} by calculating the ratio
$p_1+p_2$ to $p_{equil}$ that a unification of the
nonequilibrated fluids would yield\footnote{$p_{equil}$ and $n_{equil}$ are
calculated from eqs.\ (\ref{untangle}) by inserting $E=E_1+E_2$,
$\vec{M}=\vec{M}_1+\vec{M}_2$ and $R=R_1+R_2$ for the left-hand side.}.
The same analysis is done for the rest frame baryon densities,
\be
\langle \frac{p_1+p_2}{p_{equil}} \rangle =
\frac{\int \!{\rm d}^3x \; \frac{p_1(x)+p_2(x)}{p_{equil}(x)}}
{\int \!{\rm d}^3x}\quad,\quad
\langle \frac{n_1+n_2}{n_{equil}} \rangle =
\frac{\int \!{\rm d}^3x \; \frac{n_1(x)+n_2(x)}{n_{equil}(x)}}
{\int \!{\rm d}^3x}\quad.
\ee
The volume integration runs in one case only over a thin central layer
perpendicular to the beam axis, so that most of the matter traverses 
this plane during the reaction and in the second case over the whole
overlap region of fluids $1$ and $2$.
The ratio is by definition equal to unity in the case of unified fluids, for 
in that case $p_1$ and $p_2$ just denote the partial pressures.
One observes that kinetic equilibrium is established only after a time
$t^{eq}_{CM}\approx(5-6)~fm/c\simeq2R_{Au}/\gamma_{CM}$. A similar
equilibration time scale was also extracted from a cascade
model~\cite{GSIJB97}. Thus, observables
which are sensitive to this early stage of the reaction~\cite{Sorge}
(as e.g.\ the nucleon flow discussed in the next section) will deviate
considerably from their values in the one-fluid limit.
One also sees (cf.\ dashed curve) that the merging of projectile 
and target nucleons starts at $t_{CM}\approx5~fm/c$, when the 
collision terms $F^\mu_{1,2}$ have almost established kinetic equilibrium.

It seems that the matter in the central plane equilibrates faster than
the whole overlap region.
However one has to recall that matter is passing
through this thin layer, so that the ratios evaluated at different times
refer to different parts of matter.
Moreover, the respective matter corresponds to matter trailing the
fronts of the nuclei. On the other hand, for the overlap region
more or less the same matter remains within the volume of integration 
at all times.

\subsection{Nucleon flow}
\begin{figure}[hbp]
\hspace{1cm}{\hbox
{\psfig{figure=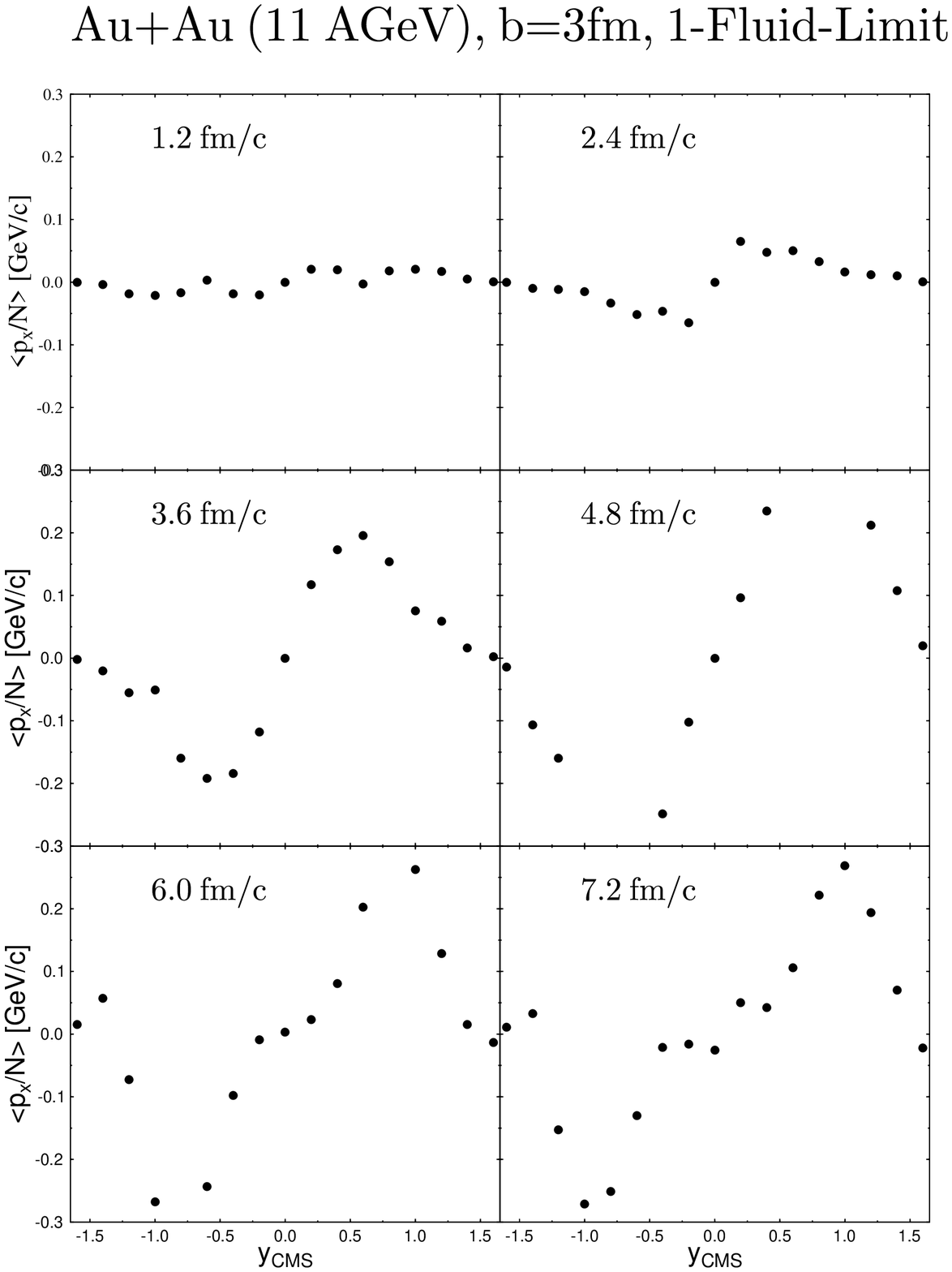,height=20cm,width=13cm}}}
\vspace*{-2cm}
\caption{The evolution of $\langle p_x/N\rangle$ in the one-fluid limit.}
\label{pxy1fluidpic}
\vspace*{.5cm}
\end{figure}

\begin{figure}[hbp]
\hspace{1cm}{\hbox
{\psfig{figure=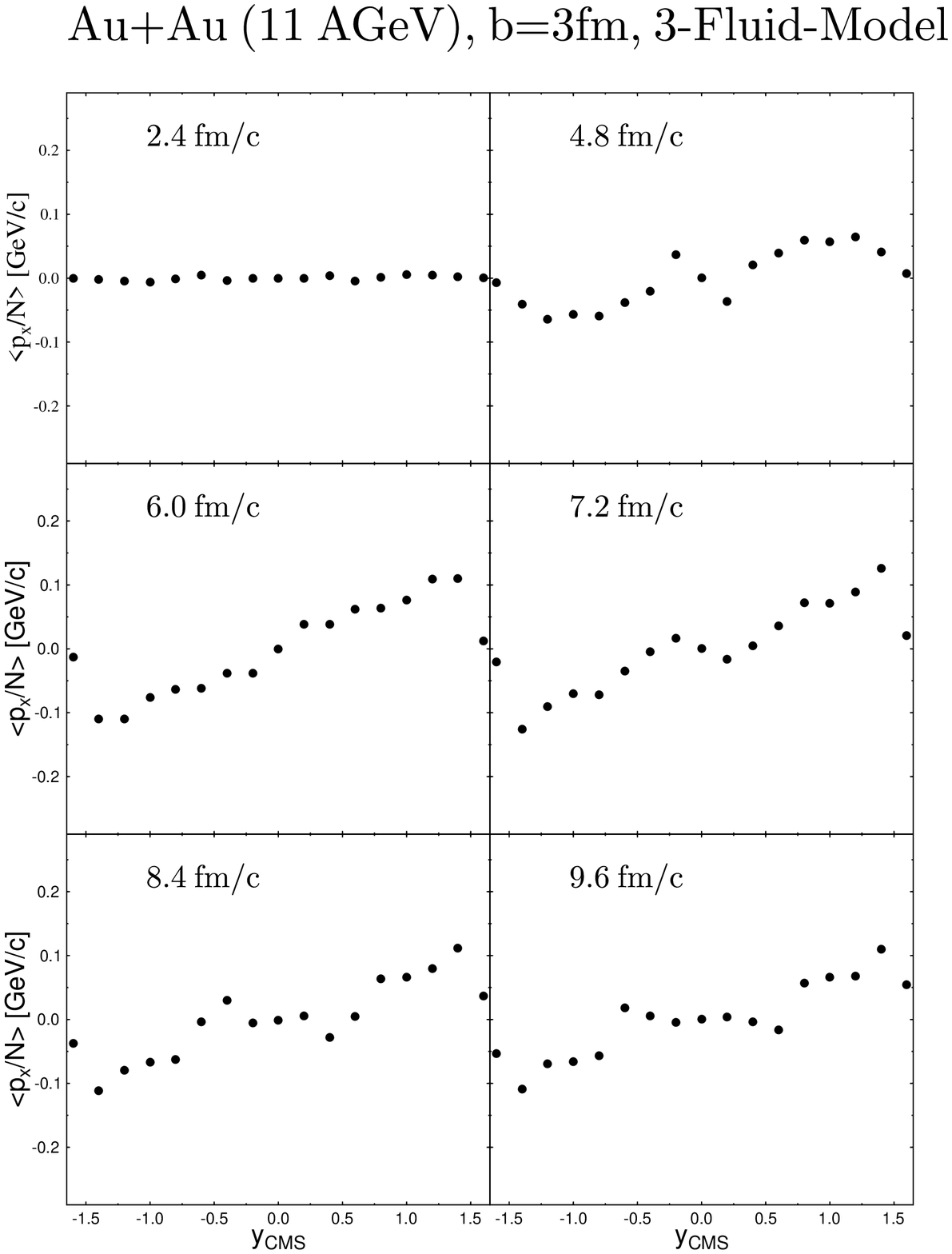,height=20cm,width=13cm}}}
\vspace*{-2cm}
\caption{The evolution of $\langle p_x/N\rangle$ in the three-fluid model.}
\label{pxy3fluidpic}
\vspace*{.5cm}
\end{figure}

As a measure for the in-plane transverse flow we investigate the mean transverse
momentum per nucleon $\left<p_x/N\right>$ as a function of rapidity $y$.
We define $\left<p_x/N\right>$ as follows:
\be \label{pxN}
\left<p_x/N\right>(y) \quad = \quad \frac{\int_y \!{\rm d}^3x \, 
R(\vec{x})\,m_N\,u_x(\vec{x})}
{\int_y \!{\rm d}^3x \, R(\vec{x})}
\ee
Here $u_x=\gamma v_x$ denotes the component of the local 4-velocity field 
in the direction of the impact parameter and $m_N$ the nucleon rest mass. 
We assume $m_N\,u_x$ to be a measure for the component of the nucleonic 
momentum in $x$-direction without thermal smearing.
The volume integration is performed over all fluid elements 
(projectile and target) around a given rapidity $y$. 
Figures \ref{pxy1fluidpic} and \ref{pxy3fluidpic} show the time evolution 
of the $\left<p_x/N\right>$-distributions for one-fluid limit and
three-fluid model, respectively.

During the reaction pressure builds up and ``spectator'' matter is deflected. 
Therefore, $\left<p_x/N\right>$ is increasing.
The one-fluid limit shows higher maximum $\left<p_x/N\right>$
than the three-fluid model, i.e.\ produces more sideward flow,
as claimed in section \ref{motiv}. In the later stages the flow is
still $\approx 280MeV/c$ in the one-fluid limit, 
which is higher than the final value of the
three-fluid model of $\approx 120MeV/c$.
The flow for the three-fluid model does not change any more from
about $6.0~fm/c$ on and thus is a reasonable measure for the final 
amount of flow after freeze-out.

In \cite{Yaris} an excitation function of the directed flow  
$\left< p_x^{dir} \right>$ has been investigated for a hadron gas EoS in 
comparison to an EoS with a phase transition to QGP.
This one-fluid model predicted a minimum of the directed flow
$\left< p_x^{dir} \right>$ around the phase transition region.
$\left< p_x^{dir} \right>$ was defined as~\cite{Yaris}:
\be \label{pxdir}
\left< p_x^{dir} \right> = \frac{1}{N} \;
\int \!{\rm d}y\; \frac{{\rm d}N}{{\rm d}y}\, 
\frac{y}{\left|y\right|} \left(\left<p_x/N\right>\left( y \right) \right) \quad.
\ee
This is basically the weighted mean of the function $\left<p_x/N\right>(y)$,
 eq.~(\ref{pxN}).

For our EoS ~(\ref{idnucleos}), $\left< p_x^{dir} \right>=120~MeV/c$ in 
the one-fluid limit, whereas $\left< p_x^{dir} \right>=40~MeV/c$ 
in the three-fluid model.
Even with our crude EoS without a phase transition~(\ref{idnucleos})
the directed nucleon flow is as low as in the one-fluid model with
a phase transition to QGP~\cite{Yaris}.
Obviously, the effect of the nonequilibrium effects is of the same
order of magnitude as the softening of the EoS due to a (first order)
phase transition.
Some implications of kinetic nonequilibrium in the early stage of
ultrarelativistic heavy-ion collisions were already sketched in \cite{Sorge}.

\section{Summary, conclusion and outlook}
In this paper we presented a three-fluid hydrodynamical model which
allows to account for nonequilibrium effects between target, projectile,
and produced particles during the early stage of the collision.
We discussed that due to the nonvanishing thermalization time scale,
this model yields a
\begin{enumerate}
\item lower energy density and temperature,
\item lower pressure,
\item less baryonic compression,
\item and a different velocity distribution
\end{enumerate}
of the nucleons at early times as compared to the one-fluid hydrodynamical 
model (which assumes {\em instantaneous} local thermalization between 
projectile, target, and produced particles). As a consequence, we expect that
the
\begin{enumerate}
\item directed nucleon flow,
\item entropy production (pion multiplicity),
\item thermal electromagnetic radiation~\cite{Dum95a},
\item lifetime of the hot and dense central region
\end{enumerate}
differ considerably in the three-fluid as compared to the one-fluid model. 
These results suggest that the predictions of the one-fluid model 
(as summarized in the introduction) have to be modified by taking 
nonequilibrium effects into account.

We especially point out that in three-fluid model the projectile and 
target nucleons reach kinetic equilibrium only after 
$t_{CM}^{eq}\approx2R/\gamma_{CM}$.
Thus the directed nucleon flow (emerging at early times) is smaller and in
particular {\em much less sensitive to the EoS} as compared to the one-fluid
limit. This is due to the fact that the flow of the nucleons of fluid $i$
is caused mainly by the partial pressure $p_i$ and not by the full 
$p_{eq}$ (see Fig.\ \ref{noneq_eqpic}).

In the future more realistic equations of state for
the projectile and target fluids (incorporating a phase transition to
a quark-gluon plasma at high energy densities) have to be implemented and
the excitation functions for the pion multiplicity, the mean transverse
momentum, and the directed nucleon flow have to be reinvestigated. 

\vspace*{1cm}
{\bf Acknowledgements:}
We gratefully acknowledge helpful discussions with L.\ Satarov and
I.\ Mishustin.
The work of D.H.R.\ has been supported by the
Director, Office of Energy
Research, Division of Nuclear Physics of the Office of 
High Energy and Nuclear Physics of the
U.S.\ Department of Energy under contract no.\ DE-FG-02-96ER-40945.

\end{document}